\documentclass[aps,prd,reprint,groupedaddress,onecolumn,showpacs]{revtex4-1}
\usepackage{color}
\usepackage{graphicx}
\usepackage{bm}
\usepackage{amssymb,amsmath,amsfonts, mathrsfs}

\usepackage{url}
\begin{document}
\begin{flushright}USTC-ICTS-16-18
\end{flushright}

\title {Partial Wave Decomposition in  Friedrichs Model With
Self-interacting Continua}
\author{Zhiguang Xiao}
\email[]{xiaozg@ustc.edu.cn}
\affiliation{Interdisciplinary Center for Theoretical Study, University of Science
and Technology of China, Hefei, Anhui 230026, China}

\author{Zhi-Yong Zhou}
\email[]{zhouzhy@seu.edu.cn}
\affiliation{School of Physics, Southeast University, Nanjing 211189,
P.~R.~China}
\affiliation{
Kavli Institute for Theoretical Physics China, CAS, Beijing 100190, China}


\date{\today}

\begin{abstract}
We consider the nonrelativistic model of coupling bare discrete states with
 continuum states in which the continuum states can have
interactions among themselves. By partial-wave decomposition
and constraint to the conserved angular momentum eigenstates, the
model can be reduced to  Friedrichs-like model with additional interactions
between the continua. If a kind of factorizable form factor is chosen, the
model can be exactly solvable, that is, the generalized discrete eigenstates
including bound states, virtual states, and resonances,  can all be represented using the original bare
states, and so do the
in-state and out-state. The exact $S$ matrix is thus obtained. We then discuss the
behaviors of the dynamically generated $S$-wave and $P$-wave discrete
states as the coupling is varying when there is only one self-interacting
bare continuum state. We find that even when the
potential is repulsive there could also be resonances and virtual
states.  In the $P$-wave cases with attractive  interaction, we find that when there is a near-threshold bound state,
there will always be an accompanying virtual state and we also give a
more general argument of this effect.
\end{abstract}


\maketitle

\section{Introduction}
Resonance phenomena appear in most areas in modern physics, such as
in optics, atomic physics, condensed matter, and particle physics.
Especially, more and more resonances were found in hadron physics,
from low energy light $\sigma$, $\kappa$ resonances to heavy quarkonium-like
resonances found in recent years. Historically, the theoretical understanding of the
resonant state dates back to the description of the nuclear alpha decay
by Gamow using eigenstates with complex energy eigenvalues, which is
also called Gamow state. The Gamow state can
not be represented as a vector in the Hilbert space since it is a
generalized eigenstate of the full selfadjoint Hamiltonian with a complex
eigenvalue. The mathematical description of the Gamow state needs an
extension of the Hilbert space
 to the rigged Hilbert space, which is composed of a Gel'fand
triple $\Omega\subset \mathscr H \subset \Omega^\times$, where
$\mathscr H$ is the usual Hilbert space of the normalizable states,
$\Omega$ is a nuclear space which is dense in $\mathscr H$, and
$\Omega^\times$ is the space of the anti-linear continuous functionals
on the nuclear space. The Gamow state should be in the  larger $\Omega^\times$.
The descriptions of in-state and out-state are
using different Rigged Hilbert spaces, $\Omega_{\pm}\subset \mathscr H
\subset \Omega_\pm^\times$ where the subscript ``$-$" denotes the out-state
space and $+$ denote the in-state space.
For further detailed discussion
on the mathematical foundation, the readers are referred
to~\cite{Bohm:1989,Gadella:2004,Gelfand4}.

Friedrichs model\cite{Friedrichs:1948} is a solvable model which
demonstrates the generation of the Gamow state, in which a discrete
bare state is coupled to a continuum state. When the energy of the
discrete state is above the threshold of the continuum state, the
discrete state will move to the second Riemann sheet of complex energy
plane of
$S$ matrix and becomes a resonant state, i.e., a Gamow state, whose
wave function can be solved. In fact, this model also demonstrates the
transition between bound states, virtual states, and resonances when the
couplings are changed. The in-states and the out-states wave
function as the energy eigenstates of the full Hamiltonian can also be
obtained.
In nonrelativistic theory, the model can either be formulated in
three-dimensional momentum space such as in \cite{Prigogine:1991} or
in a special partial-wave channel with only one continuum. However,
even in nonrelativistic theories, including spins of the continuum
states, there could be different continua with different spin
configurations and the same total angular momentum that can enter the
interactions. We will see later that, from the full Hamiltonian in
momentum representation of the Hamiltonian, after partial wave
decomposition, the Hamiltonian will be reduced to the multi-continuum
Friedrichs-like model. In fact, including more than one continua
without the interaction between the continua, the Friedrichs model can
also be solved~\cite{Xiao:2016wbs}.
The Friedrichs model
can also produce the dynamically generated states which do not
originate from the discrete state~\cite{Likhoded:1997,Xiao:2016dsx}.
Thus, Friedrichs model
provide a model to describe the observed bound states or resonances,
in which the origin of the state could be investigated.

In recent years, more and more heavy quarkonium-like states and
possible exotic states  were observed in the experiments, such as
$X(3872)$, $D_{s0}^{*}(2317)$, $Z_c$'s, and $Z_b$'s~\cite{Olive:2016xmw}, which can not be explained satisfactorily by the
conventional ``quenched" quark model such as the well-known Godfrey-Isgur
model~\cite{Godfrey:1985xj}. Efforts are made to understand the
possible mechanisms of generating these states. Take the enigmatic
$X(3872)$ for example. The $X(3872)$ state can be regarded as being generated
by coupling a bare $\chi_{c1}(2P)$ state, the state   in the ``quenched"
potential model,  to the $\bar{D}D^*$ continuum in a unified mechanism in which other charmonium-like states above the
open-flavor thresholds can also be
described~\cite{Zhou:2013ada,Danilkin:2010cc,Coito:2012vf}.
 This picture is also supported by a refined analysis of
$B\rightarrow KJ/\psi\pi^+\pi^-$ and $B\rightarrow KD\bar{D}^*$~\cite{Meng:2014ota}.
Alternatively, in \cite{Aceti:2012dd,Gamermann:2009uq}, a model was
proposed to study relation of the wave function of resonance states
and the scattering amplitude, and the method was used to discuss  $X(3872)$.
The model contains no bare discrete state, and
has only the continuum interactions and the form factor is assumed to be
factorized. The $S$ matrix and the resonance or bound state  wave
function  was obtained
by solving Lippmann-Schwinger equation in momentum representation
following~\cite{Weinberg:1962hj,Weinberg:1965zz}. In Ref. \cite{Sekihara:2014kya}, the authors
generalized this method to including also the bare discrete
states, but only $S$-wave processes are considered.  In fact, we will
show that after the partial wave decomposition, this  model can be reduced to  generalized Friedrichs-like model which
includes one or more discrete states and also   interactions
between continuum states.  If the partial wave form factor in this model can be separated to
two factors, like in \cite{Yamaguchi:1954mp,Yamaguchi:1954zz,Hernandez:1984zzb,Aceti:2012dd, Sekihara:2014kya}, this kind of generalized Friedrichs-like model can also
be solved, that is, the eigenstates, including  in-states, out-states, and the
discrete eigenstates of the full partial wave Hamiltonian can be obtained by
directly solving the eigenstate equation.
Hence the exact partial-wave $S$ matrix in this model can be obtained in this way.
Thus, all the partial waves can be dealt with in a similar fashion,
the only differences are in the form factors which should be
postulated
in different models. The discussions on the compositeness and
elementariness in \cite{Weinberg:1962hj,Weinberg:1965zz,Sekihara:2014kya} can be generalized to different
partial waves.

As examples, we also studied the behavior of the dynamically generated
$S$-wave and $P$-wave states as the coupling varies using a kind of
exponential form factor in the Friedrichs-like model with only one
self-interacting continuum state and without any discrete bare state.
If this potential is repulsive, there could still be
resonances in the $S$-wave channel and virtual states in the $P$-wave
channel. For the attractive potential, in the $S$-wave channel, there
could be a bound state for large coupling and as the coupling becomes
weaker, it will become a virtual state. In the $P$-wave, the attractive
potential will generate both a bound state and a virtual state for
large coupling, and as the coupling is turning down, the two states
move through the threshold and become a pair of  resonant state poles. As the
coupling tends to 0, the poles all move to the negative infinity.

The paper is organized as follows: In Sect. \ref{sect:nonrel}, the
partial wave analysis in the non-relativistic model is performed and
the most general model is discussed.
In
Sect. \ref{sect:Solu}, the solution to the generalized Friedrichs-like
model with a kind of factorizable form factors is given. The wave
function of the discrete eigenstates,  in-states, and out-states are
obtained and thus the partial-wave $S$ matrix is obtained. In Sect.
\ref{sect:Dyna}, we will discuss the dynamically generated discrete
states using an example form factor.  Sect. \ref{sect:conclude} is the
conclusion and discussion.

\section{Nonrelativistic Partial Wave Decomposition\label {sect:nonrel}}

In this section we will make clear the connection between the momentum
space and the partial wave expansion of the Hamiltonian, and show that
the Hamiltonian in terms of the angular momentum eigenstates will be
reduced to the Friedrichs-like models.

Suppose a discrete state $|0;ll_z\rangle$ with spin $l$,
 coupled with a continuum composed of two-particle state $|\vec
p;SS_z\rangle$ with the c.m. momentum $\vec p$ for
each particle and  total spin $S$. In the non-relativistic theory, the free Hamiltonian in the
c.m. frame  can be expressed as
\begin{align}
H_0= M_0 \sum_{l_z}|0;ll_z\rangle\langle0; ll_z| +\sum_{S_z} \int \mathrm d^3 p\, \omega
|\vec p;SS_z\rangle
\langle \vec p;SS_z| \,,
\end{align}
where $M_0$ is the rest energy
of the discrete state and $\omega=M_{th}+\frac {p^2}{2\mu}$ is the
energy of the continuum state in the c.m. frame, $M_{th}$
being the threshold energy of the two-particle states and  $\mu$ being
the reduced mass in the c.m. frame. The
normalizations and completeness relations for these states are
\begin{align}
&\langle 0;ll_z'|0;ll_z\rangle =\delta_{l_zl_z'}\,, \quad \langle \vec
p';SS_z| \vec
p;SS'_z\rangle=\delta^3(\vec p'-\vec p)\delta_{S_zS_z'}\,,
\\
&\sum_{S_z}\int \mathrm d^3p |\vec p;S_z\rangle \langle \vec
p;S_z|+\sum_{l_z}|0;ll_z\rangle\langle 0;ll_z|
=\mathbf 1
\end{align}
The plane wave state $|\vec p;SS_z\rangle$ can be decomposed into
different  partial waves
\begin{align}
|\vec p;SS_z\rangle= \sum_{LM_L}i^LY_L^{M_L*}(\hat
p)| p;LM_L,SS_z\rangle=\sum_{JM,LM_L}i^LY_L^{M_L*}(\hat
p) C^{JM}_{LM_L,SS_z}| p;JM;LS\rangle
\end{align}
where $LM_L$ in $| p;LM_L,SS_z\rangle$ denotes orbital angular momentum
quantum numbers,
$Y^{M_L}_{L}(\hat p)$ is the spherical harmonic function with $\hat p$ the
direction of $\vec p$,  and $p$ denotes the absolute value of the momentum
$\vec p$. $|p;JM;LS\rangle$ is the eigenstate of the total angular momentum
with $JM$ denoting the quantum numbers of the total angular momentum.
The free Hamiltonian can be expressed in the angular momentum
representation as
\begin{align}
H_0= M_0 \sum_{l_z}|0;ll_z\rangle\langle0; ll_z| +\sum_{J,M;L}
\int p^2\mathrm d p\, \omega |p;JM;LS\rangle \langle p;JM;LS|
 \,,
\end{align}

There is no direct interaction of the discrete state with itself,
i.e., $\langle 0;ll_z|V|0;ll_z'\rangle =0$, since it can
be absorbed into the rest energy term. The interaction between the discrete states and the continuum states
is spherically symmetric and
the matrix elements of the interaction potentials are supposed to be
\begin{align}
\langle 0;ll_z|V|\vec
p;SS_z\rangle=\sum_{LM_L}i^L\tilde g_{L}(p^2)C^{ll_z}_{LM_L,SS_z}Y^{M_L*}_{L}(\hat
p)
\end{align}
where
$C^{ll_z}_{LM_L,SS_z}$ is the Clebsch-Gordan coefficient.
Thus the interaction term in the Hamiltonian can be expressed as
\begin{align}
H_{01}=\sum_{S_z,l_z}\int \mathrm d^3p|0;ll_z\rangle\langle 0;ll_z|V|\vec p;S
S_z\rangle\langle \vec p;S S_z|+h.c.=&
\sum_{L,l_z}\int p^2\mathrm dp \tilde g_{L}(p^2)|0;ll_z\rangle\langle
p;ll_z;L S|+h.c.
\\=&\sum_{L,l_z}\int \mu p\mathrm d\omega \tilde
g_{L}(p^2)|0;ll_z\rangle\langle  p;ll_z;L S|+h.c.
\end{align}
Since the total angular momentum and the $z$ component are supposed to
be conserved,
there is no crossing terms between states with  different such quantum
numbers, and we can restrict to the subspace with fixed total angular
momentum $l$ and its $z$-component $l_z$.
One can redefine the state and the form factor to be
\begin{align}
|\omega,L\rangle={\sqrt {\mu p}}|p;ll_z;L S\rangle\,,\quad
|0\rangle=|0;ll_z\rangle
\,,\quad
 g_L(\omega)=\sqrt {\mu p} \tilde g_L(p^2)\,,
\label{eq:redefine}
\end{align}
and then the orthogonal condition reads
\begin{align}
\langle\omega,L|\omega',L'\rangle=\delta(\omega-\omega')\delta_{LL'}\,.
\end{align}
 The interaction Hamiltonian between the discrete state and the continuum in this subspace now becomes
\begin{align}
H_{01}=\sum_L\int d\omega  g_L(\omega)|0\rangle\langle
\omega,L|+h.c.
\end{align}
There could also be the direct interaction between the continuum
two-particle
states, which is supposed to conserve only the total angular momentum $J^2$
and $J_z$. To be more general, we consider the interaction between two
kinds of continuum with spins $S_1$ and $S_2$,
\begin{align}
\langle \vec p'S_2S_{2z}|V|\vec pS_1 S_{1z}\rangle
=&\sum_{JML_1'M'_{L1}L_2'M'_{L2}} (-i)^{L'_2}Y_{L'_2}^{M'_{L2}}(\hat
p') C^{JM*}_{L'_2M'_{L2},S_2S_{2z}}
i^{L'_1}Y_{L'_1}^{M'_{L1}*}(\hat
p) C^{JM}_{L'_1M'_{L1},S_1S_{1z}}\tilde f^{JM}_{L_2'S_2,L_1'S_1}(p'^2,p^2)
\end{align}
where  $\tilde f^{JM}_{L_2'S_2,L_1'S_1}(p'^2,p^2)=\langle p'JM;L_2'S_2|
V|p,JM;L_1'S_1\rangle$, $JM$ being the quantum numbers for the total
angular momentum and its $z$-components, and $L_1'$ and $L_2'$ being the
quantum numbers for orbital angular momenta. The function $\tilde
f_{L'_2,S_2,L'_1,S_1}^{JM}$ should be  decreasing sufficiently fast as
$p,p'\to \infty$, and behave as $p^{L_1}p'^{L_2}$ at  $p,p'\to0$ limit. We will see later that this threshold behavior is consistent  with the one for the
scattering amplitude.
Note that these momenta denote the momenta
of the free  states which are not the eigenstate of the full
Hamiltonian. Therefore, this interaction does not mean the non-conservation of
the energy. The real eigenstates of the full Hamiltonian are in-states
and out-states which asymptotically tends to the free states in the
$t\to \pm\infty$ limit when they feel no interaction. The $S$ matrix still
conserves the energy.
Now, the interaction Hamiltonian between continuum states can then be
expressed as
\begin{align}
H_{21}=&\sum_{S_{2z}S_{1z}}\int \mathrm d ^3p'\mathrm d^3 p| \vec
p'S_2S_{2z}\rangle \langle\vec
p'S_2S_{2z}|V|\vec pS_1S_{1z}\rangle\langle\vec pS_1S_{1z}| +h.c.
\\=&\sum_{JM}\sum_{L_2,L_1}\int \mu_2 p'\mathrm d \omega'\,\mu_1 p\mathrm
d \omega\,
\tilde f^{JM}_{L_2S_2,L_1S_2}(p'^2,p^2)|
p'JM;L_2S_2\rangle \langle pJM;L_1S_1| +h.c.
\end{align}
We have changed the integration variable from the momentum
to the c.m. energy $\omega$,  and $\mu_{1,2}$ are the corresponding
reduced masses.
Since the  interaction is supposed  not to mix states with different $JM$, we can
restrict in a subspace with fixed $JM$, and redefine
\begin{align}
|\omega,L_i;i\rangle=&{\sqrt {\mu_i p}}|p;JM;L_i S_i\rangle\,, \text { for
} i=1,2\,
,
\\
 f^{(21)}_{L_2L_1}(\omega',\omega)=&\sqrt {\mu p\mu' p'}
\tilde f^{JM}_{L_2S_2,L_1S_1}(p'^2,p^2)\,.
\label{eq:redefine21}
\end{align}
Then, the interaction term $H_{21}$ for $JM$ channel is recast into
\begin{align}
H^{JM}_{21}=&\sum_{L_2,L_1}\int\mathrm d \omega'\,\mathrm
d \omega\,
 f^{(21)}_{L_2L_1}(\omega',\omega)|
\omega',L_2;2\rangle \langle \omega,L_1;1| +h.c.
\end{align}
For the model with only one continuum, there is only one
self-interaction of the continuum which can be obtained just by
setting the continuum state to the one defined in (\ref{eq:redefine})
in above equation and the $1,2$ denoting different continua can be omitted. Thus,
the full Hamiltonian for the $JM$ channel can be expressed as
\begin{align}
H= M_0 |0\rangle\langle0| +\sum_{L}
\int \mathrm d \omega\, \omega |\omega,L\rangle \langle \omega,L|
+\sum_L\int d\omega  g_L(\omega)|0\rangle\langle
\omega,L|+h.c.
+\sum_{L_2,L_1}\int\mathrm d \omega'\,\mathrm
d \omega\,
 f_{L_2L_1}(\omega',\omega)|
\omega',L_2\rangle \langle \omega,L_1| +h.c.
\end{align}

One can generalize this model to more than one discrete states
$|j\rangle$, $j=1,2,\cdots$ and more
continuum states.  One can also regard the continuum states
with different $(L_i, i)$ combination as different states and label them
using sequential
integers $1,2,\dots $ and allow $M_{i,th}$ to have degenerate energies.
The general Hamiltonian can be expressed as
\begin{align}
H=&\sum_{j=1}^D M_j |j\rangle\langle j|+\sum_{i=1}^C\int_{M_{i,th}}^\infty d\omega_i \omega_i
|\omega_i;i\rangle\langle \omega_i;i|
\\&+\sum_{i_2,i_1}\int_{M_{i_1,th}}\mathrm d \omega'\,\int_{M_{i_2,th}}\mathrm
d \omega\,
 f_{i_2,i_1}(\omega',\omega)|
\omega';i_2\rangle \langle \omega;i_1| +h.c.
\\&+\sum^D_{j=1}\sum_{i=1}^C\int_{M_{i,th}} d\omega
g_{j,i}(\omega)|j\rangle\langle
\omega;i|+h.c.
\label{eq:Friedrichs-nonrel}
\end{align}
where $D$ discrete states and $C$ continuum states are assumed.
This is the most general Friedrichs-like model with interactions
among continuum states and discrete states.

\section{Solution to a special kind of Friedrichs-like model with interacting
continua\label{sect:Solu}}

For general form factors of the discrete-continuum and
continuum-continuum interactions, the Friedrichs-like model
is not solvable. However, if we take the form factors as
in~\cite{Yamaguchi:1954mp,Yamaguchi:1954zz,Hernandez:1984zzb,Aceti:2012dd,Sekihara:2014kya},
\begin{align}
g_{ij}(\omega)=u^*_{ij} f_j^*(\omega)\,,
f_{j'j}(\omega',\omega)=v_{j'j}f_{j'}(\omega')f_{j}^*(\omega)
\end{align}
in which $u_{ij}$ and $v_{j'j}$ are constants and the form factor $f_j(\omega)$ comes always with the $j$th
continuum state, the Friedrichs-like model is then solvable.
In this case, the Hamiltonian can be expressed as
\begin{align}
H=&\sum_{i=1}^D M_i|i\rangle\langle
i|+\sum_{i=1}^N \int_{a_i}^\infty \mathrm d \omega
\,\omega|\omega;i\rangle\langle \omega;i|
\\ &+\sum_{i,j=1}^C v_{ij}\Big(\int_{a_i}^\infty\mathrm d \omega
f_i(\omega)|\omega;i\rangle\Big)\Big(\int_{a_j}^\infty\mathrm d \omega
f^*_j(\omega)\langle \omega;j|\Big)
\\ &+\sum_{j=1}^D\sum_{i=1}^C \left[ u_{ji}^*|j\rangle\Big(\int_{a_i}^\infty\mathrm d \omega
f^*_i(\omega)\langle \omega;i|\Big)+ u_{ji}\Big(\int_{a_i}^\infty\mathrm d \omega
f_i(\omega)|\omega;i\rangle \Big)
\langle j|
\right]
\end{align}
where $v_{ij}=v_{ji}^*$ can be seen from the hermitian of the
Hamiltonian.

The eigenstate of the Hamiltonian with eigenvalue $E$ can be expanded using the discrete
states and the continuum states
\begin{align}
|\Psi(E)\rangle=\sum_{i=1}^D
\alpha_i(E)|i\rangle+\sum_{i=1}^C\int \mathrm d\omega
\psi_i(E,\omega)|\omega;i\rangle
\label{eq:expand}
\end{align}
From the eigenstate equation, one finds equations
\begin{align}
&(M_j-E)\alpha_j(E)+\sum_{i=1}^C u_{ji}^*\int_{a_i}^\infty
\mathrm d \omega f_i^*(\omega)\psi_i(E,\omega)=0
\\
&\sum_{j=1}^D\alpha_j
(E)u_{ji}f_i(\omega)+(\omega-E)\psi_i(E,\omega)+\sum_{j=1}^Cv_{ij}A_j(E)f_i(\omega)=0
\label{eq:eigen-s-eq}
\end{align}
where we have defined $A_j(E)=\int_{a_j}^\infty\mathrm
d\omega\, f^*_j(\omega)\psi_j(E,\omega)$.
If the eigenvalue $E\notin [a_i,\infty)$ for $i=1,\cdots, C$,
we can obtain
\begin{align}
&\alpha_j(E)=-\frac1{M_j-E}\sum_{i=1}^C u_{ji}^* A_i(E)
\label{eq:discret-eq}\\
&\sum_j V_{ij}(E) f_i(\omega) A_j(E)+(\omega-E)\psi_i(E,\omega)=0
\label{eq:cont-eq}
\end{align}
where $V_{ij}\equiv v_{ij}-\sum_{l=1}^D\Big(\frac
{u_{lj}^*u_{li}}{M_l-E}\Big)$ is defined.
Multiplying Eq. (\ref{eq:cont-eq}) with $f^*_i(\omega)$ and integrating
w.r.t. $\omega$, one finds the equation for $A_i$
\begin{align}
\sum_{i=1}^CM_{ji}A_i= \sum_{i=1}^C(\delta_{ji}+G_j(E)V_{ji})A_i=0\,,
\label{eq:eigenv-eq}
\end{align}
 where
$G_j(E)\equiv\int_{a_j}\mathrm d\omega
\frac{|f_j(\omega)|^2}{\omega-E}$,
$M_{ji}\equiv\delta_{ji}+G_j(E)V_{ji}$.
To have non-zero $A_i$ solution,
\begin{align}
\det M=\det[\delta_{ji}+G_j(E)V_{ji}(E)]=0
\label{eq:discret-eigen}
\end{align}
must be satisfied.  Thus, the solutions $\tilde E_i$ to Eq. (\ref{eq:discret-eigen})
will be the discrete eigenvalues for the full Hamiltonian and the
eigenvectors $A_i(\tilde E_k)$ at these eigenvalues can be solved with
the normalization undetermined, where the subscript $k$ denotes the
different eigenvalues. The number of solutions may be more than the
original bare discrete states. The extra states may be generated from
the singularities of the form factors~\cite{Likhoded:1997,Xiao:2016dsx} or by coupled channel effects
which cause shadow poles~\cite{Eden:1964zz,Xiao:2016wbs}.  Since $M_{ij}^*(E)=M_{ji}(E^*)$ and
$(\det M(E))^*=\det M(E^*)$, the
solution should be symmetric w.r.t the real axis as expected. Then from Eqs. (\ref{eq:discret-eq})
and (\ref{eq:cont-eq}), we obtain the discrete eigenstate
\begin{align}
|\Psi(\tilde
E_k)\rangle=&\sum_{i=1}^CA_i(\tilde E_k)\Big(-\sum_{l=1}^D\frac{u_{li}^*}{M_l-\tilde E_k}|l\rangle
+ G_i^{-1}(\tilde E_k)\int_{a_i}\mathrm d\omega
\frac{f_i(\omega)}{\omega-\tilde E_k}|\omega;i\rangle\Big)\,.
\label{eq:dis-eigenstate-r}
\end{align}
If the eigenvalue is real $E_B$ below the threshold on the first sheet, the state represents a bound state. As what
was done in~\cite{Sekihara:2014kya}, one can define elementariness $Z_l$
and compositeness $X_j$
\begin{align}
Z_l=&|\alpha_l(E_B)|^2=\frac{\sum_{ij}A_i u_{li}^*u_{lj}A^*_j}{(M_l-E_B)^2}
\\
X_i=&\int \mathrm d \omega\,|\psi_i(E_B,\omega)|^2= G_i'(E_B)\sum_{jj'}
A_jA^*_{j'}V_{ij'}^*V_{ij}
\end{align}
 and the
normalization of $A_i$ can be fixed by requiring
$\langle\Psi(E)|\Psi(E)\rangle =1$, i.e.
\begin{align}
\langle\Psi(E)|\Psi(E)\rangle =\sum_{ij}A_i^*(
E_B)V_{ij}'(E_B)A_j(E_B)+\sum_{jj'}
A_jA^*_{j'}\sum_iV_{ij'}^*V_{ij}G_i'( E_B)=\sum_l Z_l+\sum_i
X_i=1
\end{align}
where the prime in $G_i'$ and $V_{ij}'$ means the derivative.
The first term  $Z=\sum_l Z_l$ is just the total elementariness and
the second term $X=\sum_i X_i$ is just the
total compositeness defined in Ref.~\cite{Sekihara:2014kya}. Using the
eigenvalue equation (\ref{eq:eigenv-eq}),
the compositeness and elementariness can also be written as
\begin{align}
&X_i=-|A_i(E_B)|^2(G_i(E_B)^{-1})'\,,\quad
Z_l=
\sum_{ij}c^*_jG^*_jV_{ji}'(E_B)G_ic_i\,,
\end{align}
where $c_i( E_B)\equiv\sum_k V_{ik}A_k=-G_i^{-1}(E_B)A_i(E_B)$ is used which is just the same
definition as in \cite{Sekihara:2014kya}.

If the eigenvalue is not real, it should not be on the physical sheet
which is required by causality and should be symmetric with respect to
the real axis as explained above. The integral in
(\ref{eq:dis-eigenstate-r}) should be analytically continued to the
sheet on which the solution $\tilde E_k$ lies which can be achieved by
deforming the integral path as did in~\cite{Xiao:2016dsx, Xiao:2016wbs}.
 The
corresponding state is also represented as in
(\ref{eq:dis-eigenstate-r}). There is also the left eigenstate with the
same eigenvalue,
\begin{align}
\langle \tilde \Psi(\tilde E_k)|=\sum_{i=1}^D
\tilde\alpha_i(\tilde E_k)\langle i|+\sum_{i=1}^C\int \mathrm d\omega
\tilde\psi_i(\tilde E_k,\omega)\langle \omega;i|
\end{align}
where
\begin{align}
&\tilde \alpha_j(\tilde E_k)=-\frac1{M_j-\tilde E_k}\sum_{i=1}^C u_{ji} \tilde  A^*_i(\tilde E_k)
\label{eq:discret-eq-l}\\
&\tilde \psi_i(\tilde E_k,\omega)=-\frac 1{\omega-\tilde E_k}\sum_j \tilde A^*_j(\tilde E_k) V_{ji}(\tilde E_k) f^*_i(\omega)
\,,
\label{eq:cont-eq-l}
\end{align}
and the normalization can be fixed by
\begin{align}
\langle\tilde \Psi(\tilde E_k)|\Psi(\tilde E_k)\rangle =&\sum_{ij}A_i^*(
\tilde E^*_k)(V_{ij}'-(G_i^{-1}(\tilde E_k))'\delta_{ij})A_j(\tilde E_k)
\\=&\sum_{ij}
c_i(\tilde E^*_k)^*(G^*_i(\tilde E_k^*)V'_{ij}G_j(\tilde E_k)+\delta_{ij}G_i'(\tilde E_k))c_j=\sum_l Z_l+\sum_i
X_i=1\,.
\end{align}
Here, $Z_l$ and $X_i$ are not real any more, and thus can not have
probability explanation. However, the author of
\cite{Sekihara:2014kya} also propose them to denote the compositeness
and the elementariness of the resonance. In \cite{Guo:2015daa},
another way to describe the elementariness and compositeness of
resonances was proposed in which the two quantities can be real.

We now come to the continuum eigenstates of the full Hamiltonian.
There are $C$ continuum eigenstates for the free Hamiltonian,
and we expect that there are also $C$ continuum eigenstates which reduce to
the free eigenstates as the couplings are turned off.
The $k$th eigenstates can still be expanded using the free states as in
Eq. (\ref{eq:expand})
\begin{align}
|\Psi^{(k)}(E)\rangle=\sum_{i=1}^D
\alpha^{(k)}_i(E)|i\rangle+\sum_{i=1}^C\int \mathrm d\omega
\psi^{(k)}_i(E,\omega)|\omega;i\rangle
\label{eq:expand-cont}
\end{align}
 and equations similar to Eq. (\ref{eq:eigen-s-eq}) can also be
obtained, with superscript $(k)$ added to $\alpha$ and $\psi_i$
\begin{align}
&(M_j-E)\alpha^{(k)}_j(E)+\sum_{i=1}^C u_{ji}^*\int_{a_i}^\infty
\mathrm d \omega f_i^*(\omega)\psi^{(k)}_i(E,\omega)=0
\\
&\sum_{j=1}^D\alpha^{(k)}_j
(E)u_{ji}f_i(\omega)+(\omega-E)\psi^{(k)}_i(E,\omega)+\sum_{j=1}^Cv_{ij}A^{(k)}_i(E)f_i(\omega)=0\label{eq:eigen-s-eq-cont}
\end{align}
The continuum eigenvalue corresponding to the $k$th continuum
lies above the $k$th threshold, i.e. $x>a_k$ and is real. Since the
state reduces to the $k$th state when the couplings are turned off, there
should be a delta function in the $\psi_i^{(k)}$
\begin{align} \alpha^{(k)}_j(E)=&-\frac1{M_j-E}\sum_{i=1}^C u_{ji}^* A^{(k)}_i(E)
\\
\psi^{(k)}_{\pm,i}(E,\omega)=&-\frac 1{\omega-E\pm i\epsilon}\sum_j V_{ij}(E) f_i(\omega)
A^{(k)}_j(E)+\delta_{ik}\gamma_k(E)\delta(E-\omega)
\end{align}
Thus, from the second equation we have
\begin{align}
\sum_j M_{\pm,ij}(E) A^{(k)}_j(E)=\sum_j(\delta_{ij}+G_{\pm,i}(E)V_{ij}(E))A^{(k)}_j(E)=\delta_{ik}f_k^*(E)\gamma_k(E)
\end{align}
where $G_{\pm,j}\equiv-\int_{a_j}\mathrm d\omega
\frac{|f_j(\omega)|^2}{E-\omega\pm i\epsilon}$. We can define a matrix
$G_{\pm}=diag\{G_{\pm,1},G_{\pm,2},\cdots,G_{\pm,C}\}$, and then in
matrix form, $M_{\pm}=1+G_{\pm}V$.
Then $A^{(k)}_j(E)$ can be solved
\begin{align}
A^{(k)}_j(E)=(M_{\pm}^{-1})_{jk}f^*_k(E)\gamma_k(E)
\end{align}
( no sum for $k$) and $\psi_{\pm,i}^{(k)}(E,\omega)$ and $\alpha^{(k)}(E)$ can be obtained
\begin{align}
\psi_{\pm,i}^{(k)}(E,\omega)=&\gamma_k\Big[ \frac{f_i(\omega)f_k^*(E)}{E-\omega\pm
i\epsilon}
\sum_jV_{ij}(E)(M_{\pm}^{-1})_{jk}+\delta_{ik}\delta(E-\omega)\Big]\,,
\\
\alpha_\pm^{(k)}(E)=&-\frac{\gamma_k(E)f_k^*(E)}{M_l-E}\sum_{i=1}^C
u_{li}^*(M_{\pm}^{-1})_{ik}\,.
\end{align}
Thus the continuum state can be expressed as
\begin{align}
|\Psi_{\pm}^{(k)}(E)\rangle=\gamma_k(E)\Big[|E;k\rangle-f_k^*(E)\sum_{j=1}^C(M_{\pm}^{-1})_{jk}\Big( -\sum_{i=1}^C
V_{ij}\int\mathrm
d\omega\frac{f_i(\omega)}{E-\omega\pm
i\epsilon}|\omega;i\rangle+\sum_{l=1}^D\frac{u_{lj}^*}{M_l-E}|l\rangle\Big)\Big]
\end{align}
$|\Psi^{(k)}_+(E)\rangle$ is the in-state, and $|\Psi^{(k)}_-(E)\rangle$ is the
out-state.
If $\gamma_k(E)=1$, the continuum states can be normalized as $\langle
\Psi_\pm^{(j)}(E')|\Psi_\pm^{(k)}(E)\rangle =\delta_{jk}\delta(E-E')$.
The partial-wave $S$ matrix can then be obtained by the inner product of the
in-state and the out-state,
\begin{align}
S_{k',k}(E',E)=\langle \Psi^{(k')}_{-}(E')|\Psi^{(k)}_+(E)\rangle=
\delta(E'-E)-2\pi i\delta(E'-E)f_{k'}(E)f^*_{k}(E)(V^{-1}+G_+)^{-1}_{k'k}
\end{align}
The threshold behavior of the partial-wave amplitude is correct due
to our requirement of the threshold behavior of the form factors.
The overall $\delta$ function means the energy conservation. It is
also easy to check that the $S$ matrix is unitary.
If there is no discrete states, the $S$ matrix reduces to the one
discussed in~\cite{Aceti:2012dd}.

As an example, if  there is only one discrete state, $D=1$, and no interaction between
the continuum $v_{ij}=0$, we define $g_i=u_{1i}$ and  $V_{ij}(E)=-\frac
{g_ig_j^*}{M_1-E}$ and then
\begin{align}
(M_\pm^{-1}(E))_{ij}=(1+G_\pm(E)V(E))^{-1}=\delta_{ij}-\frac{G_ig_ig_j^*}{\eta_\pm(E)}\,,
\end{align}
 where  $\eta_\pm(E)=E-M_1+\sum_{i=1}^C |g_i|^2 G_{\pm,i}$. Thus
\begin{align}
\sum_j V_{ij}(E)(M_\pm^{-1})_{jk}=\frac{g_ig_k^*}{\eta_{\pm}(E)}\,,\quad
\sum_j (M_\pm^{-1})_{jk}\frac {g_j^*}{M_1-E}=-\frac{g_k^*}{\eta_{\pm}(E)}
\end{align}
The continuum states and the discrete states can be reduced to
\begin{align}
|\Psi_{\pm}^{(k)}(E)\rangle=&|E;k\rangle+\frac
{g^*_kf_k^*(E)}{\eta_\pm(E)}\left(|1\rangle+\sum_i g_i\int_{a_i}\mathrm d
\omega\,\frac{f_i(\omega)}{E-\omega\pm i\epsilon}|\omega;i\rangle
\right)\,,
\\
|\Psi(\tilde
E_k)\rangle=&-\sum_{i=1}^C\frac{A_ig_i^*}{M_1-\tilde E_k}|1\rangle-\sum_i\frac{A_i}{G_ig_i}g_i
\int_{a_i}^\infty\mathrm d \omega\,\frac {f_i(\omega)}{\omega-\tilde
E_k}|\omega;i\rangle\,.
\end{align}
The eigenvalues of the discrete states $\tilde E_k$ are determined by
\begin{align}
\det M=-\frac{\eta(\tilde E_k)}{M_1-\tilde E_k}=0\,,
\end{align}
and the eigenvalue equation for $A_i$ is
\begin{align}
\sum_j(\delta_{ij}-\frac {G_i g_i g_j^*}{M_1-\tilde E_k})A_j=0\,.
\end{align}
from which  we can see that $\frac {A_i}{G_i g_i}=\sum_j \frac{g^*_j
A_j }{M_1-\tilde E_k}$
is a constant independent of $i$.
Thus, the normalized discrete state is
\begin{align}
|\Psi(\tilde E_k)\rangle=&N\left(|1\rangle-\sum_i g_i\int_{a_i}\mathrm d
\omega\frac {f_i(\omega)}{\omega-\tilde E_k}|\omega;i\rangle\right )
\\
N=&\frac 1 {\eta'(\tilde E_k)^{1/2}}
\end{align}
These results are the same as was obtained in~\cite{Xiao:2016wbs}.

\section{Dynamically generated states\label{sect:Dyna}}
Another interesting case is that when there is no discrete state,
only dynamically generated states may appear, which could be resonances,
bound states, or virtual states. It is instructive to study the
different pole trajectories of this kind of states in different
partial waves as the coupling varies. The similar pole trajectory
properties when a discrete state is coupled with a continuum for
$S$-wave are studied in \cite{Hyodo2014}. A comparison of the pole
trajectories between the
cases with and without the discrete state coupling to the continuum
for higher partial waves is also studied in \cite{Hanhart:2014ssa}
using two specific models.

For simplicity, we consider only one
continuum here. The Hamiltonian is
\begin{align}
H=\int_a \mathrm d \omega\, \omega |\omega\rangle\langle \omega|
\pm\lambda^2\int_a\mathrm d\omega\int_a\mathrm d \omega' f(\omega)f^*(\omega')|\omega\rangle\langle\omega'|
\end{align}
The plus sign denotes a repulsive potential and the minus
sign an attractive potential.
The discrete states are determined by solving the equation
\begin{align}
{\bf M}_\pm(E)=\det\, M^{\pm} =1\pm\lambda^2 G(E) =1\pm \lambda^2\int_a \mathrm d\omega\frac
{|f(\omega)|^2}{\omega -E}=0
\label{eq:detM}
\end{align}
For non-relativistic theory, the form factor $f(\omega)$ should converge to zero
sufficiently fast as $\omega\to \infty$ and for a fixed partial wave
$l$, it should behave as $k^{l+1/2}\sim(\omega-a)^{(l+1/2)/2}$ as $k\to
0$. Thus, we choose an example form factor
$f(\omega)=(\omega-a)^{(l+1/2)/2}\exp\{-(\omega-a)/(2\Lambda)\}$ and choose
the reduced mass $\mu=1$ to make all quantities dimensionless. We
first look at the $S$-wave. Thus, $G$ function can be analytically
continued on the first and second sheet as
\begin{align}
G_S(E)=&\int_a \mathrm d\omega\frac
{(\omega-a)^{1/2}\exp\{-(\omega-a)/2\Lambda\}}{\omega -E}=\sqrt
{\pi\Lambda}
-e^{(a-E)/\Lambda}\pi\sqrt{a-E}(1-\mathrm{erf}(\sqrt{\frac{a-E}\Lambda}))\,,
\\
G_S^{II}(E)=&G_S(E)+2\pi i F_S^{II}(E)
\\=&\int_a \mathrm d\omega\frac
{(\omega-a)^{1/2}\exp\{-(\omega-a)/2\Lambda\}}{\omega -E}-2\pi
i(E-a)^{1/2}\exp\{-(E-a)/2\Lambda\}
\label{eq:GII-S}
\\=&\sqrt
{\pi\Lambda}
+e^{(a-E)/\Lambda}\pi\sqrt{a-E}(1-\mathrm{erf}(-\sqrt{\frac{a-E}\Lambda}))\,,
\end{align}
where $\mathrm{erf}(z)=\frac2{\sqrt\pi}\int_0^z e^{-t^2}\mathrm dt$,
$F_S(x)=|f(x)|^2$, and
$F_S^{II}(x)=-(\omega-a)^{1/2}\exp\{-(\omega-a)/2\Lambda\}$ is the
analytic continuation of $F_S$ to the
second sheet.

\begin{figure}
\includegraphics[height=4cm]{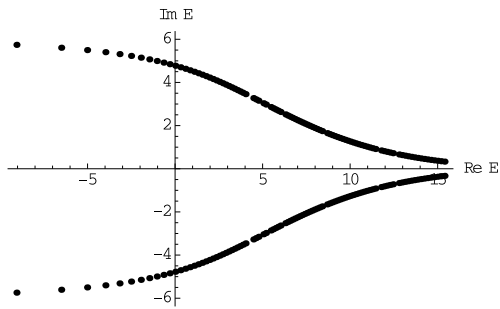}
\caption{Dynamically generated $S$-wave resonance poles on the second Riemann
sheet. $a=0.5$, $\Lambda=2$. The poles move from  right  to left
when coupling is decreasing.\label{fig:MpII}}
\end{figure}

From Eq. (\ref{eq:detM}), we can see that $\mathbf{M}_+=0$ can not
have solutions on the first sheet, since the integral is either
complex on the complex plane or positive below the threshold on the
real axis. So there could be no bound state for
this case. However, there may be resonances or virtual states on the second
sheet. From (\ref{eq:GII-S}), since the phase of $(E-a)^{1/2}$ at
$E<a$ is $\pi/2$, and the second term is positive, the equation could
not have solution for $E<a$. By numerical experiments, we find that there is a pair of
resonance poles on the second sheet. As the coupling is turning
down, the poles are moving towards the negative infinity on the complex
plane. See Fig. \ref{fig:MpII} for illustration. However, this
resonance is a little farther away compared with the $\Lambda$, and may not be
physically meaningful.

\begin{figure}
\includegraphics[height=5cm]{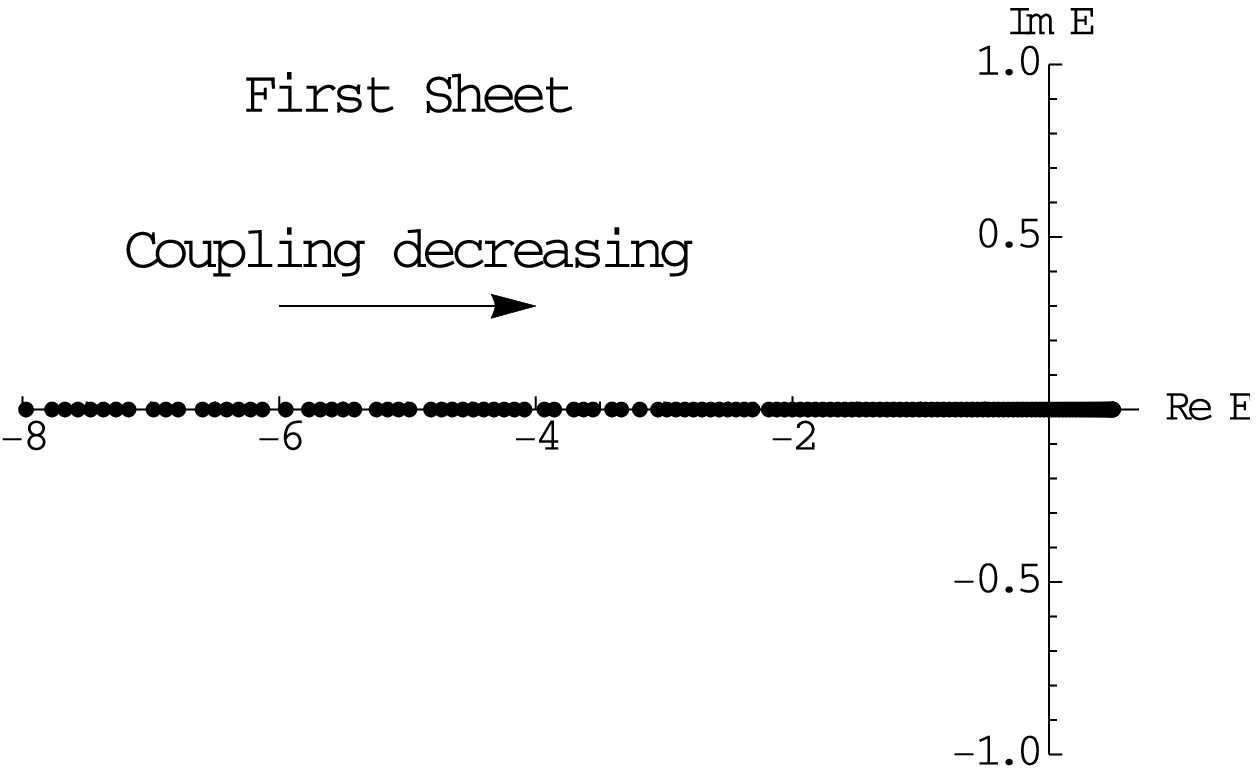}\hspace{1cm}
\includegraphics[height=5cm]{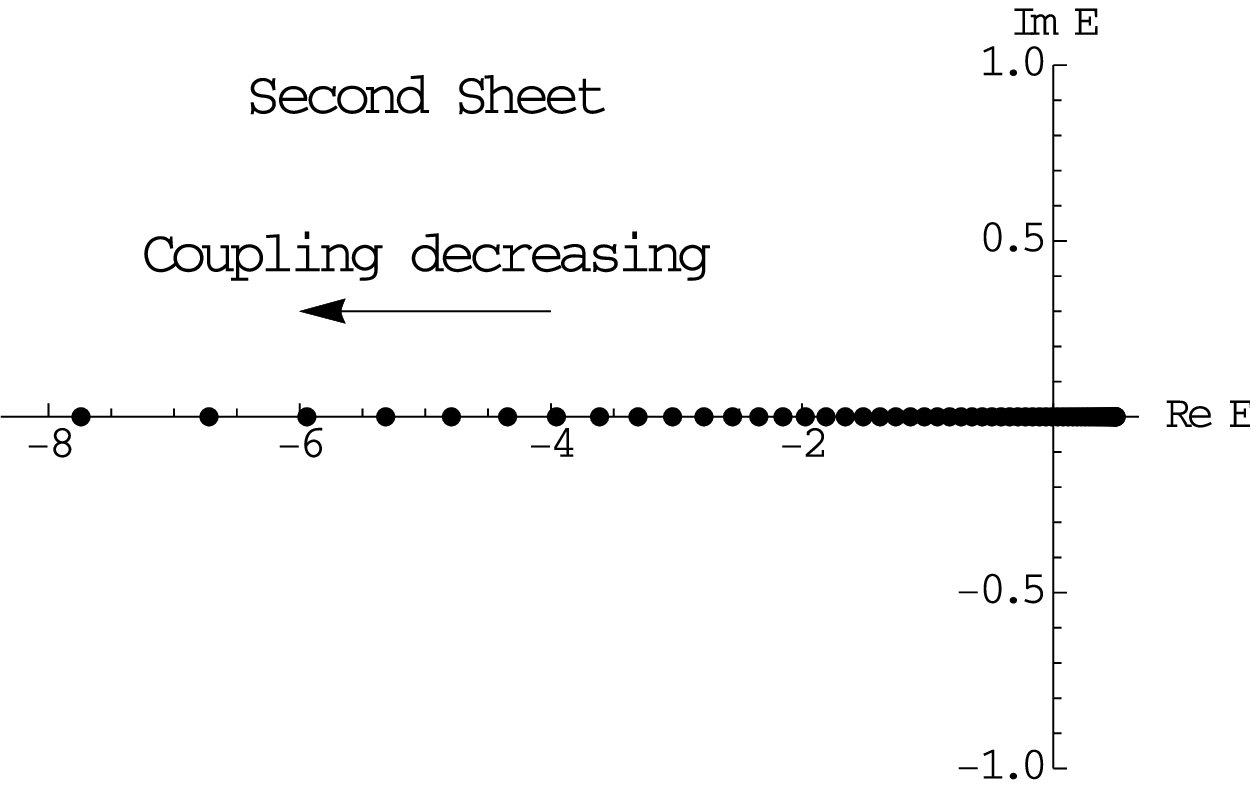}
\caption{Dynamically generated $S$-wave bound state pole on the first Riemann
sheet moves towards the threshold and then runs to the second sheet
becoming a virtual state as the coupling decreases. $a=0.5$, $\Lambda=2$. \label{fig:Mm}}
\end{figure}

However, $\mathbf M_-=0$ can have solutions on the first sheet real axis. In
fact, by numerical experiments, there is a bound state pole on the
first sheet when the coupling is large enough. As the coupling is
turning down, the bound state moves through the threshold
 to the second sheet real axis below the
threshold, becoming a virtual state. As the coupling continues turning
down further, the virtual state moves towards the negative infinity. See Fig.
\ref{fig:Mm} for an illustration. This situation is reminiscent of the
deuteron and its virtual state partner. In the nucleon-nucleon
scattering, in the spin-triplet channel there is a deuteron bound
state for a stronger coupling, while in the spin-singlet channel with
a weaker interaction, a virtual state is generated and contributes a
large scattering length.

\begin{figure}
\includegraphics[height=5cm]{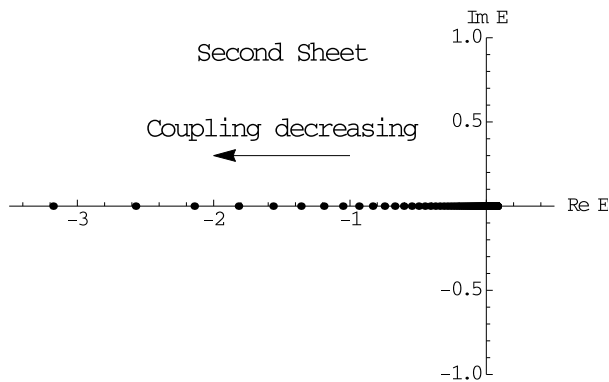}
\caption{Dynamical generated $P$-wave virtual state pole on the second Riemann
sheet moves towards negative infinity. $a=0.5$,
$\Lambda=2$. \label{fig:XIIP}}
\end{figure}

For $P$-wave, we can do the same thing. The analytically continued $G$
function on the first sheet and second sheet can be
expressed as
\begin{align}
G_P(E)=&\int_a \mathrm d\omega\frac
{(\omega-a)^{3/2}\exp\{-(\omega-a)/2\Lambda\}}{\omega -E}
\\=&\frac 12\sqrt
{\pi\Lambda}(\Lambda-2(a-E))
+e^{(a-E)/\Lambda}\pi(a-E)^{3/2}(1-\mathrm{erf}(\sqrt{\frac{a-E}\Lambda}))\,,
\\
G_P^{II}(E)=&G_P(E)+2\pi i F_P^{II}(E)
\\=&\int_a \mathrm d\omega\frac
{(\omega-a)^{3/2}\exp\{-(\omega-a)/2\Lambda\}}{\omega -E}-2\pi
i(E-a)^{3/2}\exp\{-(E-a)/2\Lambda\}
\label{eq:GII-P}
\\=&\frac 12\sqrt
{\pi\Lambda}(\Lambda-2(a-E))
+e^{(a-E)/\Lambda}\pi(a-E)^{3/2}(1-\mathrm{erf}(-\sqrt{\frac{a-E}\Lambda}))\,,
\end{align}
where $F_P(x)=|f_P(x)|^2$, and $F_P^{II}(\omega)=-(\omega-a)^{3/2}\exp\{-(\omega-a)/2\Lambda\}$ is its analytic continuation to the
second sheet.
The $\mathbf M_+=0$ still does not
have bound state solution since $G(E)$ is still positive on the first
sheet below the threshold. However, since the phase of $(E-a)^{3/2}$
term in the second term of Eq. (\ref{eq:GII-P}) is $-e^{i\pi/2}$ for
$E<a$ and the second term is negative, there is a
virtual state solution and  it will move to the negative infinity as the coupling is turning
off. This is because, the range of the first term in Eq. (\ref{eq:GII-P}) is
bounded by $(0,G_P(a))$ and  the second term is monotonically increasing
and  unbounded below. As $\lambda$ is decreasing the second term in
Eq. (\ref{eq:GII-P}) will become important. But as the coupling grows
larger, the virtual state can not go through the threshold
to the first sheet since we know that there can not be a
bound state solution for the $\mathbf M_+=0$ on the first sheet. Thus
there must be a limiting point of the virtual state as the coupling goes to
positive infinity. This point is determined by $G_P^{II}(E)=0$ which is independent of $\lambda$.
See Fig. \ref{fig:XIIP}
for an illustration.

\begin{figure}
\includegraphics[height=5cm]{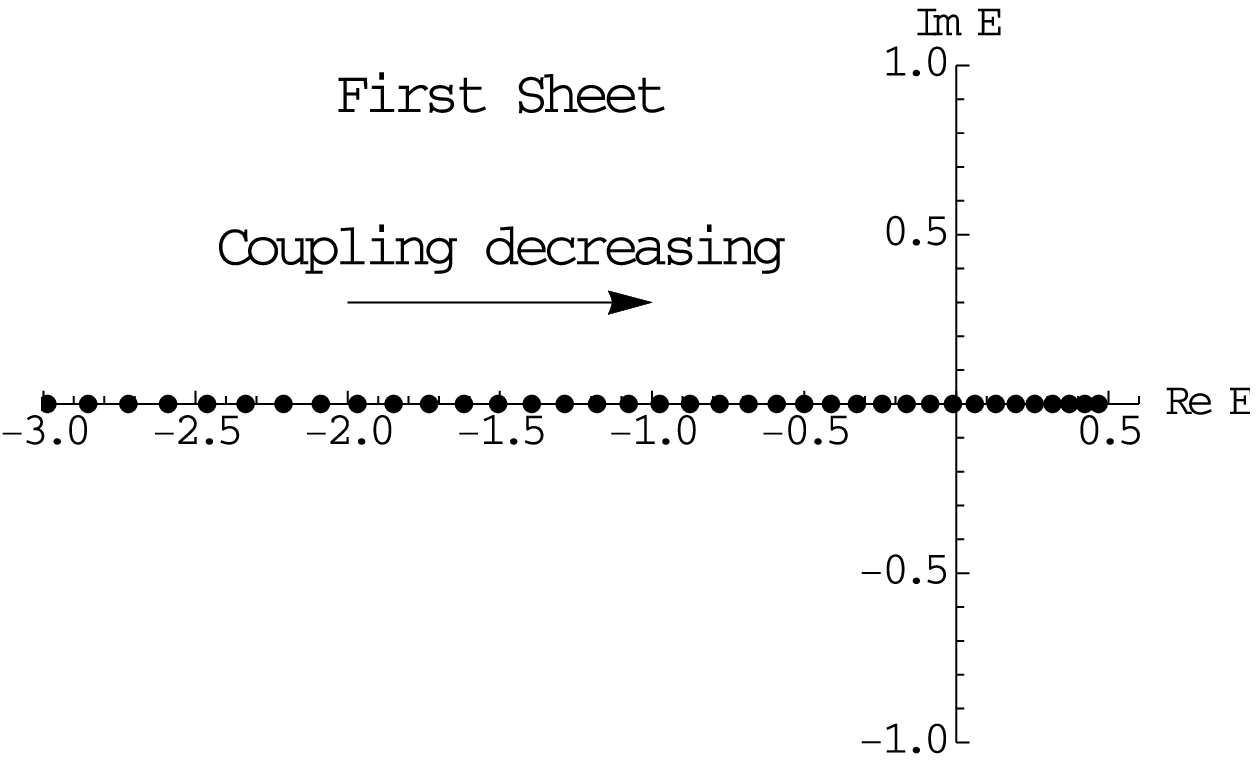}
\vspace{0.5cm}
\includegraphics[height=5cm]{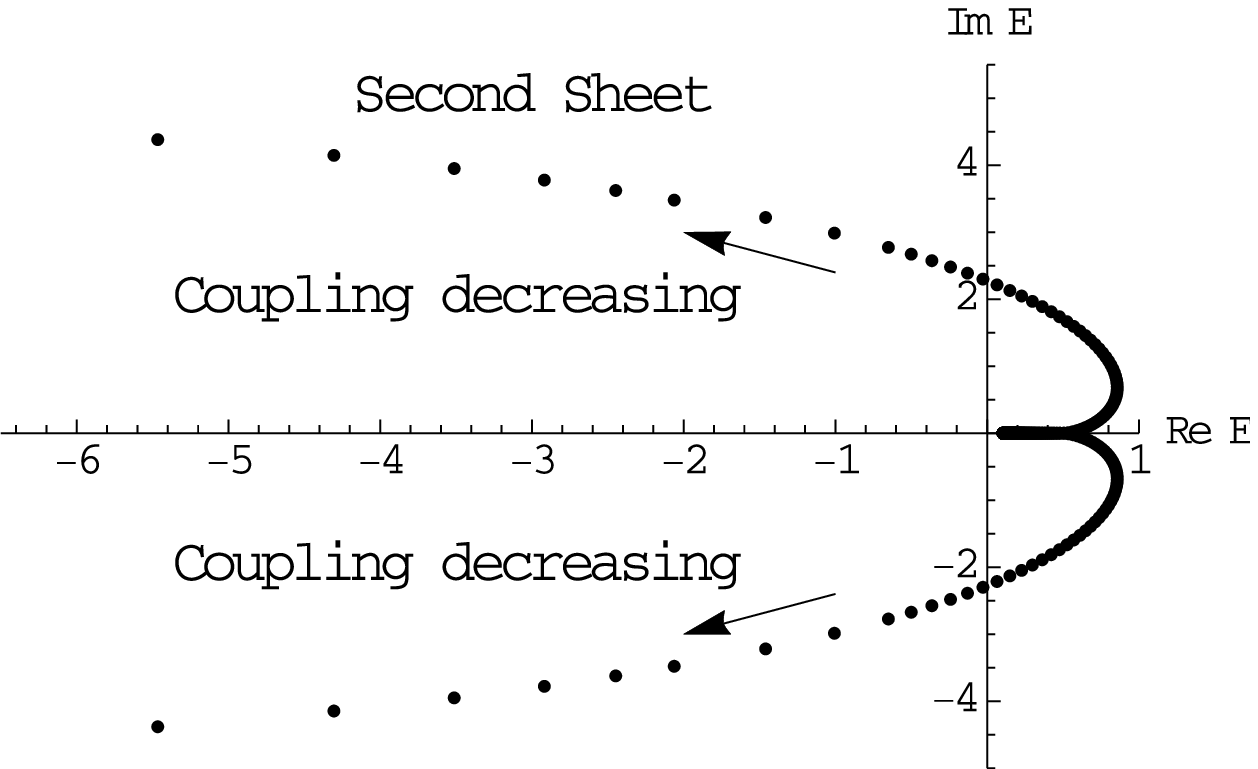}
\caption{Dynamically generated $P$-wave bound state pole on the first
Riemann sheet and virtual state pole on the second Riemann
sheet merge together at the threshold becoming a pair of resonance poles
on the second sheet as the coupling decreases. $a=0.5$,
$\Lambda=2$. \label{fig:XmP}}
\end{figure}

The solutions to $\mathbf M_-=0$ include one bound state and one virtual state for
large coupling and as the coupling decreases the two solutions are moving
through the  threshold and  becoming a pair of resonance poles on the
second sheet, and then will move towards the negative infinity. See
Fig. \ref{fig:XmP} for an illustration. If the  coupling is increased
to infinity, the bound state will move to the negative infinity, and
the virtual state will approach a limiting point on  the second sheet,
which is determined by $G_P^{II}=0$ the same as the previous case.

The existence of the accompanying virtual state of a bound state in
$P$-wave can
be understood as follows. That there is a bound state means that $\mathbf
M_-(E_B)=1-\lambda^2 G_P(E_B)=0$ where $E_B$ is the energy eigenvalue
corresponding to the bound state. Since $G_P(E)$ is a monotonically
increasing function below the threshold $a$, for $E<E_B$,
$\lambda^2 G_P(E)<1$ and for $E>E_B$, $\lambda^2 G_P(E)>1$.  On the second sheet, $\mathbf
M_-^{II}(E)=\mathbf M_-(E)-2\pi i \lambda^2 F_P^{II}(E)$, where $-2\pi
i \lambda^2 F_P^{II}(E)$ is a  monotonically decreasing function, and
for $E\le a$, $-2\pi
i \lambda^2 F_P^{II}(E)\ge 0$. Thus, $\mathbf
M_-^{II}(E_B)>0$ and $\mathbf M_-^{II}(a)<0$, and from continuity,
there must be a solution to $\mathbf M_-^{II}=0$ between $E_B$ and $a$.
The property near the threshold is determined by the
$(E-a)^{3/2}$
factor and the positivity of the form factor
$|f(\omega)|^2/(E-a)^{3/2}$ in the $F_P^{II}$. In general,
$|f(\omega)|^2/(E-a)^{3/2}$ at the $E=a$ limit should be a positive
value and would not change sign on the two sides of $E=a$, since
otherwise, $|f(\omega)|^2$ would behave according to a different power of
$E-a$ at the threshold.  So, if a dynamically generated bound state
is found to be very near the threshold
there
would also be an accompanied virtual state. In this case, it may not
be distinguished with the more fundamental bound state generated from
the discrete bare state, which is also
accompanied with a virtual state \cite{Xiao:2016dsx}. The only
difference is that in the small coupling limit the fundamental bound
state goes to the bare state, while the dynamically generated bound
state will move to the second sheet and  combines with the
virtual-state pole to form a pair of resonance
poles, and then the resonance
poles move to the negative infinity.

\section{Conclusion and outlook\label{sect:conclude}}

In this paper, we have studied the model which contains several discrete states and continuum states, in which the
interactions between discrete states and continua, and the interaction
between continua are included.
We made the partial wave decomposition of the model, and showed that
confined to a specific partial wave, it reduces to the Friedrichs-like model which include also the
interaction between continua.
If the form factors in each partial wave can be factorized as assumed
in \cite{Yamaguchi:1954mp,Yamaguchi:1954zz,Hernandez:1984zzb,Aceti:2012dd,Sekihara:2014kya}, the model can be rigorously
solved and the discrete eigenstates, the in-states, the out-states,
and the exact $S$ matrix can be obtained subsequently.

As an example, we studied the behavior of  the dynamically generated
states in non-relativistic $S$-wave and $P$-wave cases with only one
self-interacting continuum using a kind of exponential form factor.
This is a kind of typical form factor used in the phenomenological
analysis. In each case, there could be two kinds of interactions, one
with a  repulsive potential with plus sign before the interaction term
and the other with an attractive potential with a minus sign. The
$S$-wave case with a repulsive interaction has only second sheet resonances farther
away compared to the cut-off and may not be of much physical meaning.
However, for the attractive potential,
when the coupling is strong, there could be a bound state for large
couplings and when the coupling decreases, the bound state will move
through the threshold to the second sheet becoming a virtual state.
For the $P$-wave case with the repulsive potential, there is only one
virtual state and as the coupling is turning off the pole moves to the
negative infinity. For the attractive potential, there could be a
bound state and a virtual state for larger couplings, and when the
coupling is decreasing, they move to the threshold and then form a
pair of resonance poles on the second sheet. We also give an argument
that, in the $P$-wave, since the threshold behavior of the form factor
is determined by the $p^{3}$ times a positive factor, for any
potential satisfying this condition, a bound state near the threshold
will be accompanied with a virtual state. It is a requirement of the
threshold behavior.

Thus, a general nonrelativistic framework to discuss interaction
between the discrete states and the continuum is laid down. However, a
remaining problem is to generalize it to relativistic cases suitable
for particle physics application.  In fact, the essential problem for
the consistent relativistic generalization is that one must deal with
the negative frequency modes in relativistic theory. It is not easy to
include them in the Hamiltonian in the formalism used in present
paper.  There are different attempts  of relativistic generalizations
of the Friedrichs model.  One typical kind of relativistic
generalization   has been developed  by the school of Brussels,
in~\cite{Antoniou:1998JMP,Karpov:1998JMP}, in which a kind of bilocal
field is used to represent the continuum two-particle state, which is
not suitable for particle physics application. Another typical
relativistic field theory generalization is discussed
in~\cite{Antoniou:2003fk}, in which only a subset of interactions are
included in the model. Both these generalizations utilise the field
theory language and provide a clue for further work.
The other problem one must
face in applying this model in different physical situations is the
determination of the form factors. The partial-wave form factor in Eq.(\ref{eq:redefine21}) should behave as $p^{L_1+1/2}p'^{L_2+1/2}$ in the limit
of $p,p'\to0$  but also
should be converged to zero sufficiently fast as $p,p'\to \infty$ in
order for the integral to be well-defined. In different process, how
to obtain a reasonable form factor from more fundamental model such as
QCD is a challenging task. We have shown that when the form factor
can be factorized, the model can be solved. In this case, the solution to
this model is equivalent to summing over all the bubble-chain diagrams in
the field theory language, similar to the situation in \cite{Zhou:2015jta}. The
form factor in reality may not
be factorizable. Whether there could be other form factors which make the
model solvable is another research direction.

\begin{acknowledgments}
Z.X. is supported by China National Natural
Science Foundation under contract No.  11105138, 11575177 and 11235010.
\end{acknowledgments}

\bibliographystyle{apsrev4-1}

\bibliography{Ref}

\end{document}